\documentclass[manuscript,nonacm]{acmart}
\usepackage[commandnameprefix=ch,commandnameprefix=always]{changes}
\usepackage{natbib}
\usepackage{soul}
\usepackage[english,hyperpageref]{backref}

 \renewcommand*{\backrefalt}[4]{%
    \ifcase #1%
     \or (see page:~#2)%
     \else (see pages:~#2)%
    \fi%
    }

\AtBeginDocument{%
  \providecommand\BibTeX{{%
    \normalfont B\kern-0.5em{\scshape i\kern-0.25em b}\kern-0.8em\TeX}}}

\setcopyright{none}
\begin{document}

\title{Enclosed Loops: How open source communities become datasets}

\author{Madiha Zahrah Choksi}
\authornote{Both authors contributed equally to this research.}
\email{mc2376@cornell.edu}
\affiliation{%
  \institution{Cornell Tech}
  \city{New York}
  \state{New York}
  \country{USA}
  \postcode{10044}
}
\author{Ilan Mandel}
\authornotemark[1]       
\email{im334@cornell.edu}
\affiliation{%
  \institution{Cornell Tech}
  \city{New York}
  \state{New York}
  \country{USA}
  \postcode{10044}
}

\author{David Goedicke}      
\email{dg536@cornell.edu}
\affiliation{%
  \institution{Cornell Tech}
  \city{New York}
  \state{New York}
  \country{USA}
  \postcode{10044}
}

\author{Yan Shvartzshnaider}      
\email{yansh@eecs.yorku.ca}
\affiliation{%
  \institution{York University}
  \city{Toronto}
  \state{Ontario}
  \country{Canada}
  \postcode{M3J 1P3}
}


\begin{abstract}
Centralization in code hosting and package management in the 2010s created fundamental shifts in the social arrangements of open source ecosystems. In a regime of centralized open source, platform effects can both empower and detract from communities depending on underlying technical implementations and governance mechanisms. In this paper we examine Dependabot, Crater and Copilot as three nascent tools whose existence is predicated on centralized software at scale. Open source ecosystems are maintained by positive feedback loops between community members and their outputs. This mechanism is guided by community standards that foreground notions of accountability and transparency. On one hand, software at scale supports positive feedback loops of exchange among ecosystem stakeholders: community members (developers), users, and projects. On the other, software at scale becomes a commodity to be leveraged and expropriated. 

We perform a comparative analysis of attributes across the three tools and evaluate their goals, values, and norms. We investigate these feedback loops and their sociotechnical effects on open source communities. We demonstrate how the values embedded in each case study may diverge from the foundational ethos of open communities as they are motivated by, and respond to the platform effects, corporate capture, and centralization of open source infrastructure. Our analysis finds that these tools embed values that are reflective of different modes of development - some are transparent and accountable, and others are not. In doing so, certain tools may have feedback mechanisms that extend communities. Others threaten and damage communities ability to reproduce themselves. 

\end{abstract}

\maketitle

\section{Introduction}

Modern coding and development are often characterized by the assembly of complex webs of public and open-source packages. Developers typically resolve coding problems by looking at public source code or querying a search engine, technical documentation, discussions, and forums available online. From the origins of the Free Software movement to early open source to the age of the ``software supply chain'' \cite{ladisa2022taxonomy}, open source ecosystems have continuously re-invented themselves. At each stage, these changes were enabled by technical, legal, and sociological innovations codified within the open-source movement. The open-source movement produced large amounts of public source code as well as community norms around sharing, collaboration, and education. Core aspects of an open community, such as information governance, depend on established policies and common acceptable behavior. Many of the rules are derived from established contextual norms and societal expectations. As open projects scale, both the resource and the community grows, and positive feedback loops of exchange thrive~\cite{bonaccorsi2003open}. 

Starting in the early 2010's open-source ecosystems saw increasing centralization in package managers and code hosting platforms. GitHub played a central role as the home for a diverse set of open-source communities \cite{eghbal2020working, opensourcedecline}. Platform effects changed how developers write code and collaborate. Simultaneously the effect of centralizing nearly all open-source code produced new tools for interacting with software at scale. In this paper, we document three cases in Section \ref{cases} (Dependabot, Rust/Crater, Copilot) that demonstrate those interactions. Further, we examine how changes in the broader open-source community are reflected in these tools.

We outline the attributes cultivated in past open projects that maintain community norms and values, such as collaboration, transparent knowledge generation, and copyright. Past projects have carefully constructed and maintained technical and normative inroads for communities. We demonstrate that long-standing technical and normative values maintain positive feedback loops for community participation and show how modern projects can diverge and disrupt those processes. 

Further, we compare three new large-scale use of open-sourced code that operate on or within open-source projects. We have selected a set of key factors around community, knowledge production, and governing rules that we use to categorize the three case studies. We conclude with an outlook on how the change in interpretation of the aforementioned key factors might have an impact on open-source, code-sharing behavior, and community growth. 

\section{Background and Related Work}

The development and growth of open source ecosystems relied on innovations in licensing that have created complex edge cases around what constitutes free use in both the legal and ethical sense \cite{grimmelmann2015copyright}. In their work, \citet{EthicsAIOpenSource} argue for greater scrutiny of the ethical deployment of Open Source AI projects and the resulting ``downstream harms'' they might cause, which are insufficiently mitigated under current norms or legal practices \cite{cooper2023variance}.

In this section, we situate accountability and transparency in open-source movements by examining changes in the canonical values and norms operating in the broader ecosystem. Building on previous historical analysis of accountability \cite{MakingtheUnaccountableInternet}, we categorize a new mode of software at scale that raises novel questions for understanding contemporary open source systems.
This paper frames open source itself as a data commons. The mechanisms and historical provenance of datasets are a reoccurring concern in discussions of fairness, accountability, and transparency\cite{gebru2021datasheets, 10.1145/3531146.3534637, 10.1145/3531146.3533086}. Drawing on Ostrom’s Principles of Data Commons Governance~\cite{ostrom2008tragedy} to study Open Data Ecosystems (ODEs), this work extends the literature towards tools that use public source code as a material resource and commons~\cite{schweik2012internet}. The production of substantial amounts of open-source code from which these datasets are curated was predicated in legal innovations in licensing regimes. Current attempts to reign in harms from AI systems\cite{BehavioralUseLicensing} are part of a broader historical context in the effects of changes in licensing regimes on practitioners.

In the following sections, we examine the ideological, legal, and technical origins of open source. We then discuss how those factors changed in the presence of centralized platforms. This section concludes with a brief introduction to the cases we term "software at scale"  examined in the rest of this paper.

\subsection{Ideological Origins of Open Source}
The rise of the Free Software and Open Source movements in the 1980s and 90s instantiated novel norms for hosting, distributing, and accessing publicly available source code \cite{moody2009rebel}. The Free Software Foundation (FSF), and the Open Source Initiative (OSI) worked to foster and guide these movements into popular existence \cite{miller2010free}.
The enabling context for this mode of collaboration was the legal innovation of copyleft licensing, published by Richard Stallman and the FSF, known as the General Public Licence (GPL). The GPL used ``intellectual property rules to create a commons in cyberspace''\cite{Moglen_1999}. It enabled ``a commons, to which anyone may add but from which no one may subtract' \cite{Moglen_1999}. Broadly construed, copyleft licensing agreements enable copies of copyrighted work (namely programs and software) to be shared, used, and modified freely. The FSF operates as a moral crusade against proprietary software. Comparatively, the OSI focuses on the practical benefits of open-source software, such as enabling developers to modify and distribute code \cite{klang2005free}. Both movements have played a significant role in shaping the modern open-source community and its values. The modern open-source ecosystem is a product of innovations that are technical (code), legal (licensing), and sociological (norms of sharing). These factors are not static; rather, they are informed by and shaped by each other. 

\subsection{Linux Made Open Source Big}
Linux began as a side project in 1991 by Linus Torvalds to build a free operating system kernel. Its rapid popularity and its organizational mythology, as captured in Raymond's \textit{The Cathedral and the Bazaar}, helped to propel the growth of open source~\cite{raymond1999cathedral}. Its characteristics would come to define open-source development. Features include a large group of strangers arranged largely non-hierarchically while voluntarily collaborating on software over the internet. Working collaboratively in public became the basis of open source code as a ``knowledge commons''~ \cite{schweik2012internet, boldosova2015looking}. 

The key innovation of Linux:
\begin{quote}
    was not technical but sociological. Until the Linux development, everyone believed that any software as complex as an operating system had to be developed in a carefully coordinated way by a relatively small, tightly-knit group of people....

    Linux evolved in a completely different way. From nearly the beginning, it was rather casually hacked on by huge numbers of volunteers coordinating only through the Internet. Quality was maintained not by rigid standards or autocracy but by the naively simple strategy of releasing every week and getting feedback from hundreds of users
\end{quote}

When Linux switched to the GPLv2 license, it was adopted ``from the FSF, but [we believed] in it as an \textit{engineering} choice and as a way to allow people to improve and share rather than as a moral imperative.''  (\textit{Linus Torvalds, 2016} \cite{linusinterview2016}, emphasis theirs). Unlike GNU and BSD, which had a core team of developers who were physically proximate and could act as the ``inside group'' guiding development, Linux was truly made by strangers online. Linux needed the GPL, and the licensing regime informed the technical and sociological modes of the community's development. 

It is difficult to measure the totality of Linux's success in the intervening decades. Its integration in Android makes it the most popular OS in the world \cite{haris2018evolution}. Further, Linux is used in virtually every supercomputer \cite{bader2021linux} and nearly all cloud providers and web servers \cite{bernstein2014containers}. As early as 1993, it was understood that Linux was more than a project but a methodology \cite{moody2009rebel}. That methodology grew rapidly, setting the stage for a developer culture dominated by open-source software.

\subsection{GitHub Made Open Source Centralized} \label{centralization}
Originally, compressed tar files with source code were shared via FTP and mailing lists, where they would typically be compiled on user's machines. Around 1993 some distributions of Linux developed package formats to simplify installing pre-built binaries that could be installed using package managers built into the OS \cite{moody2009rebel}. These tools allowed users to install, upgrade and remove isolated packages. Debian released apt-get in 1998, making it possible to download a package and all of its dependencies \cite{claes2015historical}. This change meant code could be developed by compositing smaller, more modular components\cite{murdock2007package}. Language-specific package managers had a similar effect, changing ``the idea of what might constitute an ``open-source project'' [which] became smaller, too, not unlike the shift from blog posts to tweets'' \cite{eghbal2020working}. 

As the number of open source projects and communities expanded, there were pressures to develop organizational infrastructures on top of not just code but community as well. Git was released by Linus Torvalds in 2005 to manage the kind of decentralized collaboration that the Linux project had pioneered. In 2008 GitHub was founded as a ``Social Coding'' platform with community features layered on top of Git. GitHub emerged in the same era as YouTube, Facebook, and Twitter, as the rest of Web 2.0 was taking off. The platform integrated pull requests on top of Git, allowing developers to submit, review and merge changes to open-source code on the platform.
By 2018 GitHub announced 100 million repositories\cite{warner2019thank}. In January of 2023 they had 100 million registered users and an explicit vision to make GitHub ``the home for all developers'' \cite{dohmke2023}. Centralization has the effect of making a commons accessible as a resource \cite{illich1983silence}.

\textit{Working in Public} documents the changing nature of open source development in the late 2010s as GitHub became the de facto home for open source communities~\cite{eghbal2020working}. \citet{eghbal2020working} describes ``Federations'' and ``Stadiums'' as two possible structures for open source communities. The former is defined by high contributor growth, high user growth, and complex governance structures. Linux is a prototypical federation. Stadiums are characterized by low contributor growth and high user growth, typically powered by one or a few developers. Stadiums may have large numbers of casual contributors~\cite{casual} whose connection to a project more closely resembles that of users or the parasocial relationship between creators and viewers on platforms such as Twitch and YouTube are stadiums. Stadiums also typify many of the smaller, modular libraries that are legally reused. The organization of open communities within ``stadiums'', therefore, demonstrates a shift in community norms where developers identify as users of projects, instead of participatory and contributing members of open-source communities \cite{eghbal2020working}. 

\subsection{Case Studies} \label{cases}
The selected tools are made possible through large amounts of accessible code shared on centralized platforms and tightly integrated package managers. The cases reveal the divergent strategies of the modern open-source ecosystem as it conforms to extant platform effects. The case studies capture a novel mode of interacting with public software that is enabled by centralization while contrasting their embedded values and visions of open software development. 



\subsubsection{Dependabot}
Dependabot is a tool that maintains source code dependencies and mitigates security vulnerabilities in the software supply chain~\cite{dependabot2023}. Founded in 2017, Dependabot was acquired by GitHub in 2019, and has since been fully integrated into GitHub's platform~\cite{he2022automating}. Further, Dependabot alerts are on by default for public repositories on GitHub \cite{dependabot2023, dependabotblog}. Security Advisories are synchronized from the National Vulnerability Database\cite{nvd}, and repository owners can also raise security vulnerabilities in their code. If an advisory falls within one of GitHub's supported ecosystems, the advisory gets verified by the GitHub Security Team. While other dependency management bots exist, Dependabot's integration on GitHub makes it the most widely used. In 2019, 67\% of bot-created pull requests came from the original and GitHub native versions of Dependabot \cite{wyrich2021bots}.

\subsubsection{Crater}
The rust programming language makes an aggressive commitment to stability across updates to the language and compiler~\cite{ruststable, rfc1122}. Crater, originally called taskcluster-crater was introduced in 2015 as a tool to help guarantee stability by ``compiling and running tests for every crate on crates.io (and a few on GitHub) \cite{crater}. Currently, there are 103,318 crates on \url{crates.io} (2023-01-29). Additionally, every single public repository on GitHub with a Cargo.lock file\cite{marablog} is tested. Brian Anderson developed the initial version of Crater as ``a tool to run experiments across parts of the Rust ecosystem.''~\cite{crater}. Crater runs weekly or more \cite{marablog}. When new errors cause the compilation to fail, Rust's compiler team may revert changes to the compiler.

\subsubsection{Copilot}
Copilot is a cloud-based ``artificial intelligence'' assistant for writing code. Copilot is installed as a plugin compatible with a number of code editors, where it acts as a form of advanced auto-complete. It is currently based on OpenAI's Codex model\cite{chen2021evaluating}. Codex is a fully trained GPT3 model fine-tuned on 54 million public software repositories hosted on GitHub, filtered down to 159GB of Python source code\cite{chen2021evaluating}. It is unclear what data the deployed model was fine-tuned on, likely much more. More than 1.2 million developers joined Copilot's free technical preview in 2021. Within 1 month of moving to a \$10/month subscription model, 400,000 users signed up\cite{microsoft22}.

\section{Comparative Analysis}
The following subsections compare the aforementioned case studies across a number of attribute that create incentives for the community and maintain positive feedback loops. The goal of this analysis is to investigate how each tool is produced by platform effects in open source.

\subsection{Project Goals and Centralization}
Each tool explicitly lays out its own functional goals in documentation or marketing. In addition to these technical or operational objectives, each tool has implicit goals that define how it interacts with the broader open source ecosystem. 

\subsubsection*{Dependabot}
Within open source ecosystems, common wisdom suggests that by keeping individual packages up to date and security holes patched, the ecosystem as a whole is made safer \cite{okafor2022sok}. Dependabot attempts to keep users' code secure by helping keep everyone's code secure and up to date.
 
\subsubsection*{Crater}
In describing all the ways the Rust language is tested to maintain stability and safety one maintainer described how the existence and centralization of Github and Crates.io ``allow us to treat the entire world of open source Rust code as our test suite'' \cite{rustested}. Those tests assist Rust language developers in maintaining the state of the compiler while keeping updates to the language timely. By allocating time and compute toward such a goal, they are demonstrating their commitment to stability\cite{ruststable} for all users of the language, within the open source and outside of it.
\subsubsection*{Copilot}
Copilot's stated goal is to help developers focus on ``bigger'' problems, leaving the tedious parts of coding to an automated system \cite{CopilotAnnounce}. Copilot's unstated goal is also to be a subscription-based product for GitHub to derive a profit \cite{microsoft22}. Like nearly all social platforms from the 2010s, the free hosting of content is a means financial gain; this is no different for GitHub. Furthermore, Copilot collects telemetry data as discussed in section~\ref{UI}. This serves the other two goals by acting as a data flywheel \cite{larsson2021ai, xia2023conversational} to improve future versions of Copilot making the tool more attractive as a product.

\subsection{Platforms and Technical Implementations}
All three case studies input data would not exist without the changes in the open source ecosystem, described in Section \ref{centralization}. We examine how each tool leverages centralization and platform effects to provision data as a resource. Further, we examine how the technological material itself reflects embedded values\cite{winner2017artifacts}.
\subsubsection*{Dependabot}
GitHub scans user code and crosschecks it against a shared database of CVEs \cite{mann1999towards}. IN addition, GitHub is a numbering authority capable of adding vulnerabilities to that shared database.
Dependabot is primarily an interface to that database. The code for the bot is open-sourced but individuals cannot add to the database. They can only make suggestions that are reviewed by the GitHub security team.
\subsubsection*{Crater}
The rust language centralizes its own package management creating a data source for Crater to run on. Additional packages can be sourced from GitHub because of its dominance over code hosting. Crater currently runs on an AWS c5.2xlarge with 2 terabytes of storage\cite{crater}. This costs \$0.34/hour with runs taking a few days. The servers used by crater are restricted to certain rust team members, however the costs are nominal and individuals could presumably implement their own system.
\subsubsection*{Copilot}
Without the centralization of GitHub, collecting the massive datasets necessary for training a Codex-like model would likely be much more difficult. GitHub already hosts the code, it is technologically trivial for them to use it as they wish. Additionally, if data-set size is to be the bottleneck in improvements in transformer-like models\cite{hoffmann2022training} it is necessary to ingest maximal amounts of source code regardless of licensing or authors preferences. Training large models is financially costly, while those numbers are not released based on \citet{allal2023santacoder} it could be ~\$40-\$60,000 if compute is priced similarly to AWS EC2 P3 instances. There would be significant costs to developing and running the infrastructure for running inference as well. The tools existence and underlying technology cannot be cleaved from the social dimensions of its development \cite{whittaker2021steep}. Copilot and similar models are fundamentally products of large well funded institutions. 

\subsection{Transparency and Relationship to Open Source Licences}
\subsubsection*{Dependabot}
Released under the prosperity public license, dependabot's source code is available online as is the database alerts are based on. The tool is self-activating and on by default for public repositories. The license creates strong limitations for commercial use by a offering a restricted trial for such uses. Dependabot does not create copyright infringement or risks as its operational function to maintain the sustainability of a given repository. Dependabot's technical processes would be classified as transformational use in that its use enhances the copyrighted code towards the goal of protecting packages from security risks.

\subsubsection*{Crater}
Crater is licensed under both the MIT and Apache 2.0 licenses. The interests of the programmers who publish Rust code to crates.io and GitHub don't have expressive interests that could be violated by its reuse. Because these are public-facing repositories, regression testing doesn't violate terms of use as defined by the licenses. Further, this interest does not produce copyright issues from a fair-use perspective as it does not reduce the market value of the code that is posted. Instead, the code is made more valuable by ensuring that it continues to work. The Rust maintainers make noncommercial use of packages available online, and the functional processes Crater carries out are transformative. From a community perspective, there are no normative compelling reasons why a programmer would object to their code being scraped. Testing and stability do not invade an interest in personhood, nor do they create losses to the developer. They do not create unearned benefits to other parties. Rather, 
testing using code at scale benefits the community of which the programmer is part. It doesn't interfere with their authorship interest in the expression in the code; it doesn't interfere with their incentives to create it.

\subsubsection*{Copilot}
Whereas licenses enable others to obtain rights to copy and use software, terms of service agreements are a legal contract that outline a service relationship between a customer and a corporation. The Copilot case presents a copyright conundrum: a code recommendation system that generates copyright infringing works. However, there can be no secondary copyright liability without primary infringement, which means that Copilot is not itself liable unless it is used to infringe. Under U.S. copyright law, there are no copyrightable interests in infringing works. Therefore, Copilot as a recommendation tool that copies code produces a significant source of risk to its users. Further, Copilot is trained on code from repositories on GitHub, released under numerous open source licences but provides no attribution information. Arguably, contradicting its own normative standards for sharing code. GitHub itself admits that the text of the GPL was contained in the training data 700,000 times \cite{Ziegler_2022}. GitHub executives also argue that the tool satisfies fair use because the output belongs to the operator. 

Ongoing litigation will determine the legality of  laundering open source to sell as a subscription. Regardless of legality, a number of projects have left GitHub for at least violating the spirit of their license and their trust. One complaint asks why the models treat open source code as a free commodity but were not also trained on the proprietary code of GitHub and its owner Microsoft \cite{Kuhn_2022} if there are no copyright concerns with the system. 

\subsection{Interactions and Feedback Loops}
In this section,  we discuss how users interact with the tool and how much autonomy users have to not participate in the production of resources that enable these tools. Further, we explore how feedback loops are or are not created between users interactions, software at scale, communities, and the broader open source ecosystem.
\subsubsection*{Dependabot}
Users interact with Dependabot when alerts are raised in their repository. Those alerts are only visible to maintainers otherwise they would risk outing vulnerable projects. While alerts are on by default for public repositories, users can opt out and turn alerts off. Users can also further engage with Dependabot by raising vulnerabilities, in their own packages or others. These features are well documented on GitHub.
The safer each individual package is the safer the broader open source ecosystem is describes a feedback loop between users individual security concerns and the security of all packages.
\subsubsection*{Crater}
There is relatively little user interaction with crater by normal member's of the community. It is unclear how widespread its existence is to casual users of the programming language. While all code on \href{https://crates.io}{crates.io} is subject to crater experiments users could avoid publishing their code to the package manager. Similarly, only projects on GitHub with a \verb|Cargo.lock| file are scraped and included in Crater tests. If those files are never pushed to a public repository they will be opted out of crater runs. 
Though users may not interact with crater directly there exists a positive feedback loop between the open source ecosystem and the tool. Stability is a core commitment by the maintainers to the rust community~\cite{ruststable} and a major draw to users~\cite{zeng2019identifying} The better the Rust Foundation can keep that commitment the stronger the draw of the language, the more the ecosystem produces, providing more data for Crater to function. Even beyond the broad feedback loop, one rust maintainer described in some cases where code tested by Crater was itself unsound  `` [they] often inform the crate[package] maintainer and sometimes even help them fix it''~\cite{marablog}. 
\subsubsection*{Copilot}\label{UI}
Of the tools listed Copilot has the tightest interaction loop between users and the least transparency in a feedback loop between the community and the tool. By implementing it as a plugin for a wide variety of code editors it adds functionality to the context where developers are comfortable working\cite{barke2022grounded}. Copilot leverages textual norms of the medium of code to emphasize the ``pair programming'' modality \cite{bird2022taking}. Users report spending less time searching online for answers at the cost of understanding their own code less \cite{bird2022taking}. Github does collect telemetry data (though users can opt out) on which code suggestions have been accepted, edited, or rejected.

While these may be used to update the underlying model, feedback loops between the underlying model and the open source platform are limited temporally to updates to the underlying model. The authors of the Codex paper note that``the model may make suggestions for deprecated methods. This could increase open-source developers’ incentive to maintain backward compatibility, which could pose challenges given that open-source projects are often under-resourced''. One of the key innovations of working in public was the rapid iteration cycle that allows productive feedback loops to develop. Copilot is unlikely to be updated fast enough to produce those loops.

A significant number of users pushing Codex-like generated code to public repositories reinforces the use of popular languages and open source packages. The long term effect entrenches the disparities in package usage rates. The authors of Codex note how differential import rates might:
\begin{quote}
    increase the dominance of an already influential set of individuals and organizations in the software supply chain.
    ...
    Where a user may have done an Internet search before for 'which machine learning package to use' or 'pros and cons of PyTorch vs. Tensorflow' they might now just type '\# import machine learning package' and trust Codex to do the rest. Users might be more inclined to accept the Codex answer under the assumption that the package it suggests is the one with which Codex will be more helpful. As a result, certain players might become more entrenched in the package market and Codex might not be aware of new packages developed after the training data was originally gathered. 
\end{quote}
At a sufficient scale of users this may hinder newer projects and libraries from gaining a foothold and developing their own community. An algorithmic and computational mono-culture.




\section{Implications and Discussion}
\subsection{Incentivizing Community}
Successful open source projects, software, and communities are those that have valuable code, an active community of users, and a broader network that stimulates collaboration, dialogue, and interaction~\cite{comino2007planning}. However, these success-defining factors are not weighted equally. The quality of the resource is a culmination of the time and effort developers dedicate to make the code better~\cite{bonaccorsi2003open}. This is sometimes achieved by organizing tasks into parts and incentivizing developers to contribute in small but highly meaningful ways. And while the end product of high quality code and technical achievements are important to code being correct, maintainable, and modular, the process is more significant than the outcome~\cite{raymond1998homesteading}. 
The key to long-term success is a thriving community of developers who are motivated and incentivized to support the project solely for the social benefit that maintaining the project brings to others - regardless of their membership or activity within the community. Along these lines, our analysis investigates three attributes integral to a positive feed-back loops within an open source community: a project’s goals, values, and norms as they relate to the process.

\subsection{Values in Design}
The three case studies examined in this paper fit an emerging typology for applications of large source code data-sets. They are a fundamentally new technological object that communities will increasingly interact with, by choice or not. The values embedded in these tools and the norms they encourage will likely continue to shape open source communities for the foreseeable future. In open source ecosystems, accountability and transparency in technical systems are intimately tied to the historical arc these communities have traversed. 

It is important to acknowledge that these paths are not predetermined. Understanding if implementers of these systems perceive of the open source ecosystem as a collaborator, a commons or a resource to be extracted is predictive of downstream concerns. Of the three cases, Copilot would seem to most challenge the norms and ethos of canonical open source development. This is not predicated on technological determinism, rather is was dependant on choices made by the developers of the system. Amazon's Copilot competitor makes an effort to specify when code completions are similar to specific samples in the dataset and will provide licence and attribution information \cite{amazon_2022}. The Software Freedom Conservancy has convened a working group to determine what a truly Free and Open Source (FOSS) Copilot alternative would look like \cite{Kuhn_2022}. The BigCode project has worked to produce a large language model for code using only permissively licensed code\cite{allal2023santacoder}. They are working towards that goal using only permissively licensed code with a robust opt-out process\cite{kocetkov2022stack}.

\subsection{Long Term Viability}
The reliance on open source code or data raises the question of sustainability: For how long can new models and products rely on the open-source community to provision these goods? As the need for data increases so will the need for the provision of quality data becomes a concern~\cite{Linar2022}. Platforms will need to encourage participation to satisfy computational requirements~\cite{Linar2022}. Without supporting governance structures that lead to the production and maintenance of a digital commons, systems break due to unaddressed vulnerabilities, broken dependencies\cite{SociotechnicalRisk,9426043} and invalid datasets~\cite{curto2022sustainability}.


Regardless of the social formulations that develop these tools, they push the open source ecosystems towards a model that contains more "stadiums." This model is comprised of a large number of open source packages that are co-dependent on one another, and a large user-base who are not necessarily members of those communities. For Dependabot and Crater, these qualities are implicit in the functionality of the tool, which are predicated on user bases that create many small packages and libraries rather than monolithic code bases with a persistent community. Copilot, explicitly encourages the "stadium" model. It promotes the most popular packages in its training data-set and substitutes understanding~\cite{vaithilingam2022expectation, barke2022grounded, bird2022taking} for developer efficiency. Copilot functionally isolates developers from communities by being the first, and maybe only resource that they interact with. Popular discourse has expressed concerns about Copilot and large language models replacing developers, however, the real risk is the replacement of communities.

\section{Conclusion}

In this paper, we present an investigation into the socio-technical effects of feedback loops in open source communities. We first trace the historical and ideological origins of open source through the modern era. We examine how forces in centralizing those ecosystems produce distinct social formulations of open source communities. Similarly, centralization creates a resource of open source code as a data-set.

We classify a distinct form of ecosystem wide tools. Our comparative analysis of three open-source tools (Crater, Dependabot and Copilot) reveals how feedback loops coupled with divergence from open-source community goals, values, and norms, could hinder community formation and sustainability. 


\bibliographystyle{ACM-Reference-Format}
\bibliography{bibliography}

\end{document}